# Implementing Performance Competitive Logical Recovery

David Lomet
Microsoft Research
Redmond, WA
lomet@microsoft.com

Kostas Tzoumas
Aalborg University
Denmark
kostas@cs.aau.dk

Michael Zwilling
Microsoft Corporation
Redmond, WA
mikez@microsoft.com

## ABSTRACT

New hardware platforms, e.g. cloud, multi-core, etc., have led to a reconsideration of database system architecture. Our Deuteronomy project separates transactional functionality from data management functionality, enabling a flexible response to exploiting new platforms. This separation requires, however, that recovery is described logically. In this paper, we extend current recovery methods to work in this logical setting. While this is straightforward in principle, performance is an issue. We show how ARIES style recovery optimizations can work for logical recovery where page information is not captured on the log. In side-by-side performance experiments using a common log, we compare logical recovery with a state-of-the art ARIES style recovery implementation and show that logical redo performance can be competitive.

## 1. INTRODUCTION

### 1.1 Motivation

We believe that providing transactions to accommodate new platforms, e.g. cloud or multi-core, can be realized via a different way of architecting database systems. In particular, we want to construct database systems as a collection of modules with well-defined interfaces. This may permit vendors to assemble specialized database systems or alternatively to provide a general database system composed of or extended by multiple parts. This is not a new goal. A composable database system has been explored [3], as have extensible database systems [7].

While the limited impact of prior efforts suggests caution, the emergence of new platforms makes this a good time to reconsider database architecture. Deuteronomy proposes to decompose a database system storage engine into transactional component (TC) and data component (DC). Earlier papers have described the interface between TC and DC [10,12] and locking without location information to provide concurrency control [13]. Here we focus on recovery without the use of location information, and in particular, we demonstrate that this "logical" recovery has the ability to rapidly recover from a system crash.

Even without re-architecting a database system, logical recovery can be useful to maintain replicas at sites without a physically isomorphic environment. That is, the data can be replicated in a database using a different kind of stable storage, e.g. a disk with different page size, or flash memory with different block size. Because the log records shipped to the replica are logical, they can be applied to disparate physical system configurations.

### 1.2 Making Recovery Logical

Traditional ARIES style recovery [15] consists of two passes: the redo and the undo pass, in addition to an initial "analysis" pass that generates data structures that enable recovery performance to be optimized. The (transaction) undo pass is done in a logical way in ARIES to facilitate concurrency and deal with data movement. However, the redo pass is "physiological". While the redo operation on data in a page is expressed logically, each log record names exactly the one page affected by the operation. Such physical information cannot be present in a logical log record. While a Deuteronomy operation also updates a single page, it does not know what page. Similarly, a replica with a different hardware environment cannot exploit physical information to update its replica. Both these settings require logical recovery.

Logical redo recovery identifies records being updated using their logical attributes, e.g. table name and record key. Typically, DBMSs use the table name to locate a B-tree that is then searched using the key to find the record. Thus, for logical redo recovery, the page affected by a logged redo operation has to be discovered during recovery using the B-tree index. With current state of the art recovery methods like ARIES or like SQL Server's multi-level recovery with system transactions [4, 11], the B-tree index used for data placement is not accessed during redo recovery, but rather only during (logical) undo. Thus, SQL Server recovers the B-tree index after redo recovery and only before its logical undo recovery. The index is only guaranteed to be well-formed by the start of transactional undo recovery. For multi-level recovery, there is an extra system transaction undo pass after physiological (ARIES) redo and before the user level transaction undo pass to ensure that B-tree indexes are well formed.

Thus the recovery paradigm currently used needs to be changed for logical recovery. Logical redo recovery (undo is already logical) requires that any index used for data placement be well-formed before redo recovery can begin. This index can only be recovered at the data server, since only it knows about data placement and indexing. Thus, data server (Deuteronomy's data component or DC[1]) recovery must take place first, prior to the transactional mechanism (Deuteronomy's transactional component or TC) resending its operations to the data server for transaction level redo and undo.

---
[1] We will use DC generically to denote a data server and TC to denote a transactional mechanism.

## 1.3 Recovery Performance

Logical recovery performance is an important consideration. Recovery needs to identify the records updated by operations. These records are identified via table name and key, and not with a page identifier (PID). During redo recovery, the re-submitted operation must re-traverse the table's B-tree in order to find the page on which to redo the operation. Then, the page LSN (pLSN) is compared to the log record LSN to determine whether the operation needs to be re-executed (the idempotence test). Contrast this with traditional "physiological" recovery [6, 15]. Here, the log record contains the PID. Hence, only the last steps of comparing the pLSN to the log record LSN and redoing the operation is required. Thus, logical recovery needs extra processing to re-traverse the B-tree on every redo operation.

Extra index traversal is not the biggest problem faced by logical redo. ARIES style recovery has optimizations that greatly reduce the redo time [15]. The dirty page table (DPT), an approximation of the database (dirty) cache at the time of the crash, is used to prune the set of log records needing redo. Any page not in the DPT does not need to be read into the database cache as such a page does not need redo. In addition ARIES defines a recovery LSN (rLSN) as the earliest operation that dirtied a page in the DPT. A logged operation with an LSN < rLSN for a page in the DPT does not need redo. This can be discovered before reading the page into the database cache. Finally, the DPT helps in pre-fetching pages into the cache prior to operations needing them, reducing latency.

Without PIDs on the log for updated pages, a DPT cannot be created. In that case, all pages found during the logical redo index search need to be fetched into the database cache. Further, this fetching can only be done on demand, i.e., without using pre-fetch, which cannot be performed in the absence of page information.

So how can TC logical recovery be competitive with physiological recovery, where the PIDs are known from log records? It needs help from the DC, the component that knows the pages updated and when they are flushed. But if the DC were to log updates, we defeat the purpose for separating TC and DC, which is to partition functionality. However, having the DC log physical information that only it knows to reduce its recovery time is both sensible and is essential to minimizing down time.

## 1.4 Contributions and Outline

In Section 2, we compare ARIES style optimized physiological recovery using a DPT to unoptimized pure logical recovery. The basic logical recovery algorithm is very simple, but its performance suffers greatly in comparison with optimized ARIES. Unlike ARIES, it does not exploit a DPT which would permit it to avoid extra fetches for pages not needing recovery.

In Section 3, we describe how SQL Server builds a DPT, which is an interesting alternative to ARIES in that the DPT does not need to be captured during normal operation via checkpointing. Our first contribution, in Section 4 is to show how to provide a DPT for logical recovery. Our second contribution is to show how prefetching can also be made to work with logical recovery from the same information used to construct the DPT.

Recovery performance has rarely been quantified by experiments, and only once to our knowledge [17] in a side-by-side comparison, and that was in a main memory setting for online games. Our third contribution, in Section 5, shows how to control conditions to provide a valid side-by-side experimental comparison of logical recovery with physiological recovery, where both use exactly the same log. Our experiments are based on our logical recovery prototype derived from SQL Server 2008. The results demonstrate that logical recovery can have comparable performance with physiological recovery with a modest amount of DC logging. Section 6 presents some conclusions and related work.

The body of the paper is augmented by appendices. In Appendix B, we present a simple performance analysis that provides an explanation for our experimental results. Appendix C shows the impact of checkpoint interval on recovery performance, while Appendix D explores trade-offs between normal execution overhead and recovery performance in construction of the DPT.

## 2. COMPARING RECOVERY METHODS

### 2.1 Why Focus on Redo

Redo performance is by far the most important part of recovery performance, except perhaps for very rare cases when long running transactions need to be undone. In our experiments, the analysis pass accounts for less than 2% of recovery time, both for logical and physiological recovery. Thus we focus on redo performance. But we want to explain how redo recovery fits into the complete recovery picture, so that the reader will know what we are omitting and why.

ARIES recovery includes physiological redo and logical undo passes, these passes also performing recovery for the B-tree structure modification operations (SMOs) [15]. SQL Server increases concurrency for B-tree SMOs by using system transactions, and providing their undo recovery in a separate pass after physiological redo and before logical undo so that logical undo continues to see a well formed B-tree [4]. Logical recovery needs to move B-tree SMO redo and undo ahead of transactional redo as redo is also logical and needs a well-formed B-tree [12].

There is little performance difference between our logical recovery and prior physiological recovery except for redo. The frequency of SMOs is a property of the B-tree, not of the recovery method, and is very low compared with data updates. All recovery variants provide B-tree SMO recovery using similar log records. Hence the time spent in recovery for these operations is, to a first approximation, the same for all methods. The only difference in methods is the time at which these SMO recovery operations are executed. Further, all variants also perform logical undo as the last pass of recovery, and hence this performance is constant in all methods. Thus we focus on the performance of the redo pass. It is the largest part of recovery time with other aspects of recovery being both very similar and relatively minor factors.

### 2.2 Physiological Redo Basics

ARIES redo recovery is described as "repeating history" [15]. This may be misleading. The redo log is scanned in "history order" during redo, but not all operations are replayed. Rather, each redo log operation is tested as to whether its effects are already in stable storage. Only if not is the redo log operation re-executed. Thus, operations are re-submitted in log order and subjected to a redo test, as described in [14]. The redo test is an idempotence test. It enables operations to be submitted multiple times for "at least once" execution, tested to ensure "at most once" execution, resulting in "exactly once" execution.

In physiological recovery [6], the redo portion of a log record identifies the page changed by an update operation, the operation itself, and its argument. ARIES log records also contain a logical undo operation. The undo operation needs the potential to be "logically" executed, e.g., by re-traversing a B-tree, because the record needing undo may have been moved to another page by the time undo recovery occurs. This cannot happen during redo and hence redo recovery always knows the page, via the log record PID, for which redo is to be considered.

To determine whether re-execution of a physiological operation is needed, a page log sequence number (pLSN) is stored on each page denoting the latest operation that updated the page. If a redo operation's LSN $\leq$ pLSN of the page identified by its PID, then redo is not needed. Otherwise, the operation is re-executed, updating the page.

Even physiological redo recovery would be expensive were we to fetch into the database cache every page mentioned by a redo log record to apply a redo test. However, we can optimize the redo test, identifying log records for which the redo test can be known to fail and hence with no redo required, via the use of an ARIES style DPT. Two facts can be exploited to optimize the redo test.

1. If a page is not dirty at the time of a crash, then no redo log record operation for the page needs redo. The DPT is a conservative approximation of the set of pages dirty at the time of a crash, i.e., it guarantees that if a page is dirty at the time of crash, its PID is in the DPT.
2. Redo log operations for a page with LSNs less than the rLSN, the LSN of the operation that first dirtied the page in cache do not need redo. A conservative rLSN is included with DPT entries, conservative in that it is not greater than the LSN of the operation that dirtied the page.

SQL Server provides both these optimizations. The DPT is constructed during the recovery analysis pass, which precedes the redo pass. Enabling its construction during recovery requires some extra logging and runtime overhead during normal execution. During the redo pass following analysis, redo log records are submitted in log order. Each record is tested by a redo test to determine whether its operation needs redo. A naive redo test simply tests the pLSN on the page identified by the redo log record's PID. The ARIES redo algorithm described in Algorithm 1 performs the optimized redo test by checking for the page's entry in the DPT and its rLSNs.

---
**Algorithm 1** ARIES redo
---
1:  **procedure** ARIES-REDO-PASS(startLSN)
2:      **for** $\forall$ logRec with logRec.LSN $\geq$ startLSN **do**
3:          currLSN = logRec.LSN
4:          e = DPT.FINDENTRY(logRec.PID)
5:          **if** (e = null $\vee$
6:              currLSN < e.rLSN) **then**
7:              //Bypass current record and continue to next record
8:              **continue**
9:          //Request the page p from the buffer manager
10:         p = BPOOL.GET(PID)
11:         **if** currLSN $\leq$ p.pLSN **then**
12:             //Bypass current record and continue to next record
13:             **continue**
14:         **else**
15:             REDOOPERATION(p, logRec) //Redo the operation

The optimizations of the redo test (lines 5 and 6) are significant. Line 5 (test for null DPT entry) avoids reading in pages we never need. Line 6 (test of rLSN) delays the need to access the page, as we do not yet need it. This permits us to continue with redo recovery while the page is brought into memory.

## 2.3 Naive Logical Recovery

When we place transactional functionality in a TC and data management functionality in a DC, the TC does transactional recovery. It does not know how records map to pages at the DC, which is responsible for doing this mapping and managing the database cache. Because of this separation (information hiding), PIDs for pages being updated are not included in the TC's redo log records. Indeed, to support replicas, there may be several DCs, each storing records in a different manner, and on different pages.

During logical redo recovery (Algorithm 2), the TC submits redo operations to the DC for it to re-execute or not depending upon whether the operation's effects have already been captured in stable storage before the crash. Upon receiving the redo request, the DC searches for the record, e.g. in the table's B-tree index, based on the log record provided key. It then reads the leaf (data) page identified in the index search into its cache if it is not already present. The DC compares the LSN of the log operation with the pLSN of this page to decide whether to redo the operation or not (line 11 of Algorithm 2). With this unoptimized strategy, every page updated since the checkpoint determined redo scan start point must be brought into cache in order to perform the redo test. Some pages may be brought in more than once if memory pressure forces the cache manager to drop pages that are needed again later. This is logically correct but can greatly slow down recovery.

---
**Algorithm 2** Basic logical redo
---
1:  **procedure** TC-BASIC-LOGICAL-REDO-PASS(startLSN)
2:      **for** $\forall$ logRec with logRec.LSN $\geq$ startLSN **do**
3:          DC-BASIC-LOGICAL-REDO-OPERATION(logRec)
4:
5:  **procedure** DC-BASIC-LOGICAL-REDO-OPERATION(logRec)
6:      currLSN = logRec.LSN
7:      //Traverse the index to find the PID referred to by logRec
8:      PID = BTREE.FIND(logRec.key)
9:      //Request the page p from the buffer manager
10:     p = BPOOL.GET(PID)
11:     **if** currLSN $\leq$ p.pLSN **then**
12:         //Operation does not need to be redone
13:         **return**
14:     **else**
15:         //Redo the operation
16:         REDOOPERATION(p, logRec)

Our hypothesis is that by constructing a DPT for logical recovery, we can optimize the redo test and the rebuilding the dirty page cache, making logical redo competitive with physiological redo.

## 3. PHYSIOLOGICAL REDO WITH DPT

The DPT is an approximation of the dirty part of the buffer pool at the time of crash. As shown in Section 2.2, it is used to avoid the redo of operations whose results are already captured in stable storage. A DPT consists of entries of the form (PID, rLSN, lastLSN). PID is the page identifier of the page. rLSN is the recovery LSN, an approximation of the LSN of the first operation that dirtied the page. lastLSN is the LSN of the last operation on the page. The lastLSN is used to help construct the DPT but does not, itself, play a direct role in redo recovery.

It is not possible to construct a completely accurate DPT as knowledge of the buffer pool at the time of the crash is too costly to maintain accurately. Instead, recovery methods construct a conservative estimate that is "safe", i.e., one that contains at least the necessary dirty pages at the time of the crash. Pages in the DPT but not in need of redo are either not accessed or the pLSN redo test indicates redo is not needed. These pages are unnecessarily brought into the cache. DPT safety also requires the rLSN of each dirty page not be greater than the LSN of the first redo operation that dirtied the page. When the rLSN is too small, unnecessary rLSN test failures may occur, but correctness is ensured by the pLSN test for the page after it is brought into the cache. The tradeoff in DPT construction is between normal operation overhead and redo time. An accurate DPT minimizes redo time but needs more effort during normal operation, while a more conservative DPT requires less during normal execution but increases recovery time.

## 3.1 ARIES Checkpointing

ARIES captures the DPT during runtime. Whenever a page is updated, it is marked as "dirty" in the database cache, and given an rLSN equal to the LSN of the update. When a dirty page is written to stable storage, it is marked as "not dirty". During checkpoint a DPT is constructed including "dirty" pages in the cache, and is written as part of a checkpoint "record". The DPT is initialized during recovery with the DPT captured in the last checkpoint record. During the analysis pass, starting from the last checkpoint, every log record that references a page not in the DPT creates a new DPT entry with rLSN equal to the LSN of that log record.

## 3.2 SQL Server Checkpointing

SQL Server follows an approach that requires less runtime overhead and logging. The DPT is neither constructed nor saved during normal execution. SQL Server uses a penultimate checkpointing scheme. When a checkpoint request is issued, SQL Server writes a "begin checkpoint" (bCkpt) log record. It then starts a process that flushes dirty pages (buffers) from the cache. The result is that at the completion of the checkpoint, all pages updated by log operations that precede the bCkpt log record are known to have their results in the stable database, and hence they do not need redo. SQL Server distinguishes pages dirtied before the bCkpt log record from pages dirtied after bCkpt. It places a bit on each page buffer that is flipped when bCkpt is written. Pages dirtied (a subset of those updated) during the checkpoint thus have a different bit value from those dirtied before, and are not flushed. When this process finishes, SQL Server writes an "end checkpoint" log record (eCkpt).

During recovery, this means that only pages updated since the penultimate checkpoint (the last bCkpt record with an accompanying eCkpt record) can be dirty and recovery can begin with an empty DPT as of this point. No updates logged earlier than this bCkpt are responsible for pages dirty in the cache when the system crashed. The analysis pass starts from this bCkpt log record and adds PIDs to the DPT from log records written after the bCkpt log record. Thus the redo scan start point is at this last bCkpt log record and earlier log records are ignored.

The analysis pass adds to the DPT every update log record PID, when first encountered in the log scan. The first mention of a page sets the rLSN of the page's entry in the DPT, and every subsequent mention of the page updates its lastLSN. This results in an unnecessarily large DPT, since no knowledge about page flushes has been exploited. Indeed, that is the case for ARIES as well. In order to prune the DPT during analysis, page flushes are monitored and logged in batches during normal operation.

## 3.3 Tracking Page Flushes

When a page is flushed during normal operation, a callback to an IO completion adds the PID of the page to an array of "flushed" PIDs that is maintained by SQL Server during normal operation. In addition, if this is the first PID to be captured in the array, the end of the stable log is captured as the first-write LSN (FW-LSN). Periodically, a so-called Buffer Write log record (BW-log record) logs this array, and then empties it. Thus, the BW-log record contains the PIDs of pages flushed since the previous BW-log record, as well as the captured FW-LSN:

$$\text{BW-logRec} = (\text{WrittenSet}, \text{FW-LSN}).$$

Not every flush might be captured, which may result in a more conservative DPT and consequently increased redo time, but this does not affect correctness. In particular, any flushes following the last BW-log record written to stable storage before a system crash are not captured.

During the analysis pass, BW-log records are used to prune the DPT. Every PID in an update log record, if not already present, is added to the DPT. When a BW-log record is encountered, SQL Server removes from the DPT under construction pages with PIDs contained in the record's WrittenSet whose lastLSN $\leq$ FW-LSN. These pages were flushed after their last update, so all their updates have been captured in stable storage. In addition, remaining page entries whose rLSNs are smaller than the FW-LSN now have their rLSNs set to FW-LSN since these pages were flushed in a state that included all updates earlier than FW-LSN. Thus, the LSN of the operation that dirtied any remaining page (i.e., its rLSN) must have a lower bound of FW-LSN. The DPT is also updated during redo, when information about the pLSNs of the pages becomes available. Algorithm 3 shows how SQL Server constructs the DPT during the analysis pass.

**Algorithm 3** SQL Server DPT construction during analysis

```
1:  procedure SQL-SERVER-ANALYSIS-PASS(bCkptLSN)
2:      DPT = null
3:      for ∀ logRec with logRec.LSN > bCkptLSN do
4:          currLSN = logRec.LSN
5:          if logRec is an update log record then
6:              e = DPT.FINDENTRY(PID)
7:              if e == null then
8:                  DPT.ADDENTRY(logRec.PID, currLSN)
9:              else
10:                 e.lastLSN = currLSN
11:         else if logRec is a BW-logRec then
12:             for ∀ PID in BW-logRec.WrittenSet do
13:                 e = DPT.FINDENTRY(PID)
14:                 if e ≠ null then
15:                     if e.lastLSN ≤ BW-logRec.FW-LSN then
16:                         DPT.REMOVEENTRY(pid)
17:                     else if e.rLSN < BW-logRec.FW-LSN then
18:                         e.rLSN = BW-logRec.FW-LSN
```

## 4. LOGICAL REDO WITH DPT

To optimize its redo test, logical recovery also needs a DPT, like ARIES and SQL Server. In our setting, this is done by the DC writing relevant information to its log. As described in [12], the DC logs B-tree SMOs in any event to make the B-tree well-formed prior to TC resubmitting its redo log records. This is essential because logical operations require that storage structures, e.g. a B-tree, be re-traversed to find the page on which the operation is to be applied since logical log records do not contain PIDs.

The DPT construction algorithms require information about pages updated and pages flushed in order to build the DPT and to assign rLSNs to its pages. Only the DC knows that information. Hence, unless it passes that information to the TC (which violates the good programming practice of information hiding), it is the DC that must remember that information across system crashes. As with SMOs, it can remember the information by logging it.

Our division of labor has the TC locking and logging logically while the DC handles data access and cache management. This avoids redundant costs. The DC logs the PIDs of pages made dirty in a batched manner, and as with SQL Server, it logs the pages flushed. This is the information needed to determine the pages in the DPT and their rLSNs. Page PIDs are in physiological redo log records, but *not* in logical redo log records. So having the DC log the PIDs of dirty pages logs no extra information, it only changes who logs it. And it is the DC that exploits both the dirty set and the flushed set to improve its recovery performance by optimizing the rebuilding of its cache.

## 4.1 Δ-log records

As indicated above, the DC needs to monitor pages made dirty as well as pages made clean during normal execution. It does this in a way similar to SQL Server's technique for flushed pages. In this section, we describe a Δ-log record, written during normal execution, that contains the information that we need to optimize recovery. How this information is used to construct our DPT during recovery is described in the following section 4.2.

The TC and the DC coordinate during normal execution to prepare for recovery by means of two control operations that are in addition to data operations that the TC sends to the DC. These operations, which affect DPT construction, are:

**EOSL:** The TC regularly sends the DC an LSN (called eLSN) marking its "end of stable log". The TC guarantees that any operation with LSN ≤ eLSN is on the stable log. The DC uses this information for cache management (it can flush pages to stable storage that are dirtied only by operations with LSN ≤ eLSN), and to prepare for optimized redo recovery, as explained below. EOSL is the operation used to enforce the write-ahead log protocol.

**RSSP:** The TC controls its "redo scan start point", the point on its log at which it starts sending operations to the DC during recovery, by sending the DC an LSN (called rsspLSN). When the DC replies to this operation, it must have flushed to stable storage all pages dirtied by any operation with an LSN ≤ rsspLSN, ensuring that such operations do not need redo. Thus RSSP is the operation by which the TC performs checkpointing.

More about these operations, including how the TC implements them, is described in [10, 12].

We write the Δ-log record containing DirtySet and WrittenSet PID arrays, FW-LSN, FirstDirty, and TC-LSN to the DC log during normal execution. The Δ-log record format is thus the following:

Δ-logRec = (DirtySet, WrittenSet, FW-LSN, FirstDirty, TC-LSN).

When an update for a page occurs, its PID is appended to DirtySet. When the IO completes and is ack'd for a page flush, the flushed page PID is appended to WrittenSet. The FW-LSN (the TC "end of stable log" at the time of the first write) is recorded, similar to SQL Server. We also record a value called FirstDirty, which is the index in DirtySet of the PID for the first dirtied page after the first flush (the first updated page after the FW-LSN is written). Finally, the TC-LSN is the value of eLSN from the most recent EOSL operation, marking the TC's "end of stable log" at the time the Δ-log record is written.

When a Δ-log record is written, we reset its fields so that the monitoring can start from scratch for the interval to the next Δ-log record. Note that unlike SQL Server's BW-log records, recovery correctness requires that all dirtied pages be captured in DirtySet. If a dirty page is not captured in a Δ-log record, the DPT during redo may not contain a dirty page and redo of an operation may be falsely avoided. Figure 1 gives an overview of the elements involved with logical redo recovery and its optimization. Part (A) of the figure shows the normal operation work to prepare for recovery, including Δ-log records.

## 4.2 Logical DPT Construction

TC and DC synchronize their recovery preparation. The TC informs the DC as to when it is checkpointing its log, and what the LSN of the bCkpt record is. The TC will start its redo scan at that log record, ignoring log records earlier than that, once the DC has confirmed that it has flushed all the relevant pages. At that point, the TC writes the eCkpt record marking the checkpoint as complete. The DC knows that pages updated earlier than the bCkpt record have been made clean. Thus, the DC can start with an empty DPT at this point. The DC learns the LSN of the bCkpt record as the rsspLSN sent to it by the TC's RSSP, and it records this value on its log so that it knows which part of its (the DC's log) needs to be scanned during redo recovery.

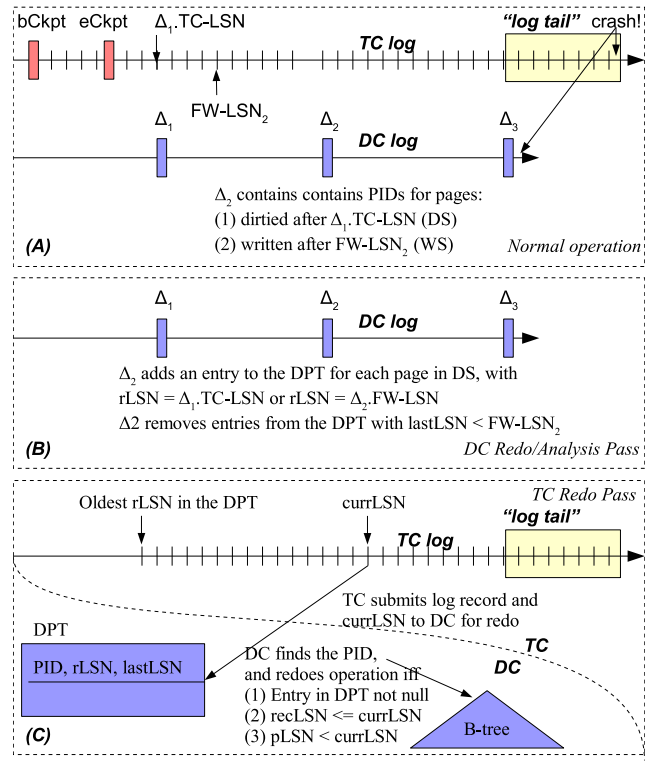

**Figure 1:** Schematic showing the operation of logical recovery with the DPT optimization. The figure includes normal operation, DC analysis pass, and TC redo pass sections.

Before the TC redo pass begins, a DC redo pass constructs the DPT using the Δ-log records only. DC recovery takes place before the TC redo recovery to make sure that B-trees are well-formed and to permit the DC to optimize TC redo recovery by constructing the DPT. Thus, DC recovery includes the optimization task that in ARIES is done by the analysis pass. Since the DC log is short (e.g. no TC redo operations), this DC redo pass processes a much smaller log than that needed for the analysis pass with integrated recovery. After DC recovery, B-trees are well-formed and the DPT has been built, with pages in the DPT having each been assigned an rLSN. Further, we have identified the TC-LSN of the last Δ-log record, and have recorded that information.

The DC rebuilds the DPT starting at the first Δ-log record with a TC-LSN greater than the last rsspLSN that the DC recorded on its log. For each Δ-log record encountered during the DC redo/analysis pass, all the PIDs in the dirty set are added to the DPT. We distinguish the PIDs that were dirtied before the first write (FW-LSN), i.e., those whose index in the DirtySet is less than FirstDirty, and those that were dirtied after this first write. For the ones dirtied earlier, their rLSN is set to the TC-LSN of the previous Δ-log record (for the first Δ-log record encountered after the RSSP, we use rsspLSN). For the latter, their rLSN is set to the FW-LSN contained in the Δ-log record. If a PID was dirtied both before and after the first write occurred, the later update will change lastLSN to FW-LSN for the updated page in the DPT. This process and the distinctions it makes is the reason for recording the FirstDirty index.

Then the flushed set of the $\Delta$-log record is used to prune the DPT. Using the flushed set, we prune DPT entries that refer to pages last updated before the FW-LSN (the eLSN at the time of the first write). These pages were either added from a previous $\Delta$-log record or they were dirtied in the current interval, but before the first write occurred (and hence in both cases were dirtied before the FW-LSN). All of these entries have a lastLSN less than the FW-LSN. Part (B) of Figure 1 illustrates the work done during the DC redo/analysis pass to optimize for TC logical recovery, and Algorithm 4 shows detailed pseudocode.

---

**Algorithm 4** DPT construction in logical recovery

1: **procedure** DC-ANALYSIS-PASS(bCkptLSN)
2:   DPT = null
3:   prev$\Delta$LSN = bCkptLSN
4:   **for** $\forall$ $\Delta$-logRec with $\Delta$-logRec.TC-LSN > bCkptLSN **do**
5:     DC-DPT-UPDATE($\Delta$-logRec, prev$\Delta$LSN)
6:     prev$\Delta$LSN = $\Delta$-logRec.TC-LSN
7:
8: **procedure** DC-DPT-UPDATE($\Delta$-logRec, prev$\Delta$LSN)
9:   $i = 0$
10:  **for** $\forall$ PID in $\Delta$-logRec.DirtySet **do**
11:    **if** $i < \Delta$-logRec.FirstDirty **then**
12:      DPT.ADDENTRY(PID, prev$\Delta$LSN)
13:    **else**
14:      DPT.ADDENTRY(PID, $\Delta$-logRec.FW-LSN)
15:    $i = i + 1$
16:  **for** $\forall$ PID in $\Delta$-logRec.WrittenSet **do**
17:    e = DPT.FINDENTRY(PID)
18:    **if** e $\neq$ null **then**
19:      **if** e.lastLSN < $\Delta$-logRec.FW-LSN **then**
20:        DPT.REMOVEENTRY(PID)
21:      **else if** e.rLSN < $\Delta$-logRec.FW-LSN **then**
22:        e.rLSN = $\Delta$-logRec.FW-LSN

---

### 4.3 Logical Redo Using the DPT

Once the DPT has been constructed by the DC, the TC redo pass begins. The TC submits logical redo operations to the DC. These are the same operations as submitted during normal execution. As in the basic logical redo (Algorithm 2), the key values in the log records are used by the DC to traverse a B-tree and identify the PIDs of updated pages. Unlike basic logical redo, the DC can now perform the redo test using the DPT it has constructed. However, it cannot do this for every log record. It works in two modes, based on whether the LSN of the TC operation is (1) less than or equal to or (2) greater than the TC-LSN of the last $\Delta$-log record.

The PIDs of pages updated after the last $\Delta$-log record (termed the "tail of the log") are not captured in the DPT. They are not available to the DC when it constructs the DPT. This information is in a DC buffer at the time of the crash, but has not been written to a $\Delta$-log record before the crash. For the tail of the log, we fall back on the basic logical redo algorithm, bringing every updated page into memory as needed and on demand. While this seems dramatically worse than SQL Server recovery, it is not. In SQL Server, records dirtied after the last BW-log record may also have an entry in the DPT unnecessarily. The set of pages needing to be read into the database cache is the same in both cases. But there is an important difference. SQL Server knows the PIDs for the dirty pages early and can, perhaps, pre-fetch them, something that is not possible for the DC since it does not know these PIDs during DC recovery. It only identifies these PIDs when it searches its B-tree for the logical operations during TC redo recovery. Part (C) of Figure 1 illustrates the work done by the DC during the TC redo pass and Algorithm 5 provides detailed pseudocode.

---

**Algorithm 5** DPT-assisted logical redo

1: **procedure** DC-LOGICAL-REDO-OPERATION(logRec)
2:   currLSN = logRec.LSN
3:   //Traverse the index to find the PID referred to by logRec
4:   PID = BTREE.FIND(logRec.key)
5:   **if** currLSN < last$\Delta$LSN **then**
6:     e = DPT.FINDENTRY(PID)
7:     **if** e = null $\vee$ currLSN < e.rLSN **then**
8:       **return**
9:   //Request the page p from the buffer manager
10:  p = BPOOL.GET(PID)
11:  **if** currLSN $\leq$ p.pLSN **then**
12:    **return**
13:  **else**
14:    REDOOPERATION(p, logRec)

---

Constructing the DPT limits the pages that need to be fetched during redo. However, when the early steps of the optimized redo test fail (i.e., the page involved is in the DPT and the log record LSN is greater than the rLSN for the page in the DPT), redo needs to wait for the page to be fetched before it can perform the pLSN test and perhaps execute the operation on the page. We cannot eliminate the "need to wait", but we can reduce the waits that we encounter by prefetching pages. We describe page pre-fetching details in Appendix A. While this does not have as dramatic an effect on redo performance as the DPT, it does result in a noticeable improvement as shown in our experiments of the next section.

### 4.4 Page Prefetch

Redo can issue a page prefetch request for a page before it encounters the log record that references this page. Thus, the page may already be available in the database buffer and redo avoids a stall. Further, by redo requesting pages in a batch, the buffer manager can group contiguous pages and read them in a block, exploiting locality of access. The DPT can assist in this. Thus, page prefetching can both reduce the number of stalls and the total number of IOs. In logical recovery, it is attractive to prefetch the internal index pages (above the leaf level) as well, since those will need to be accessed for every log record encountered. We describe index page prefetching in logical recovery in Section A.1, and data page prefetching for both logical and traditional recovery in Section A.2.

## 5. PERFORMANCE STUDY

### 5.1 Our Prototype

We implemented a logical recovery prototype derived from SQL Server 2008. Our goal was to compare logical recovery performance with the performance of physiological ARIES style recovery. For that reason, we wanted to be able to run our experiments in as controlled a way as possible. The result, as described below, is that our logical recovery experiments execute recovery using the same log with the same checkpointing as done for SQL Server executing physiological recovery.

We used a modified SQL Server database engine for normal execution. Because the recovery log must serve for both logical and physiological recovery, it must contain information that enables both forms of recovery. So, although logical log records do not use PIDs, we do not remove PIDs from the SQL Server log records, but ignore them during logical recovery. To remove them would break SQL Server recovery. Further, we log the auxiliary information needed by both recovery strategies. Thus, our normal execu-tion modifications consist of providing $\Delta$-log records for logical recovery as well as BW log-records for SQL Server recovery. This auxiliary information is a very small part of the log and does not seriously perturb either normal execution or recovery performance.

The major part of our effort was to implement logical redo recovery so that we could compare its performance with the SQL physiological redo. SQL Server uses log record PIDs to identify the changed pages. For logical redo, we discover the PID of the changed page by searching the B-tree index using the key from the log record. Optimizing redo is done in log passes prior to redo. The "DC redo" pass, which prepares the DC for optimized logical redo by building the DPT, scans the integrated log from the last checkpoint and is executed instead of the corresponding SQL Server analysis pass. This pass uses the $\Delta$-log records to build the "logical" DPT and ignores normal operation log records.

## 5.2 Experimental Setup

Our prototype implementation provides a common set of "crashes" and recovery logs to do side by side redo performance tests. In our experiments, we compare the following methods.

**Log0:** Logical redo of Algorithm 2.

**Log1:** Logical redo of Algorithm 4, with DPT, without prefetch.

**Log2:** Logical redo of Algorithm 4, with DPT, with page prefetch.

**SQL1:** SQL Server redo, with DPT, without prefetch.

**SQL2:** SQL Server recovery, with DPT, with prefetch.

All results reported are for a single-table database of two attributes, "key" and "data". The table size is approximately 3.5GB (436,000 pages, $10^8$ rows). A clustered index of 832 pages (approximately 7MB) is created on the key. The B-tree has three internal index levels. In all cases, the index could be main memory resident, a common situation for B-trees, whose fanout usually produces an index less than 1% of the size of the data. The workloads are update-only, and consist of small transactions (10 updates per transaction) that update the data attribute in a record identified by an equality search on the key attribute. Unless otherwise stated, a workload runs for double the time needed to fill the cache before we conduct our experiments so that the cache is in steady state when we do our recovery tests.

We vary the database cache from 64MB (approximately 2% of database size), to 2048MB (approximately 60%). For each cache size, we run an update workload until the cache is in equilibrium. Specifically, we crash the server when 10 checkpoints have been taken, 40000 updates have been seen since the last checkpoint, and 100 updates have been seen since the last $\Delta$/BW-log record. This causes the portion of the log that will be redone to contain approximately 40000 log records, and the tail of the log to contain approximately 100 log records in all cases. The crash happens shortly before a checkpoint is taken, which is the worst case for redo recovery. $\Delta$-log records are written exactly before BW-log records in order to ensure a fair comparison with SQL Server recovery.

In Appendix C, we examine and report on results when we vary the checkpoint interval. In the next section, we report on the impact of database cache size.

## 5.3 Some Results

Figure 2(a) shows the redo time of all the methods in milliseconds as database cache size is varied. Note that the performance of the methods, with the exception of Log0, is negatively affected by a larger cache size. Recall that the number of pages that Log0 will request (Equation 1) is equal to the length of the redo log since checkpoint (approximately 40000 log records in our case). A larger cache size can only help Log0, providing more room for these pages.

With a DPT, the number of pages requested is approximately equal to the number of DPT entries, as our analysis in Appendix B captures. Larger cache size causes the DPT to grow, and hence the redo time. However, this growth is sub-linear (the x axis in Figure 2(a) is in log scale). At every checkpoint, flush activity will cause buffers to be cleaned, giving more room for dirty pages in the cache. Since the checkpoint interval is the same for all the cache sizes, a smaller cache tends to be more "dirty" at the time of the crash. Figure 2(b) shows the DPT size as a percentage of the cache size. The DPT size varies from 30% to 10% of the database cache size. The largest cache sizes do not fill sufficiently to require early flushing in the checkpoint interval and the DPT continues to grow with unflushed pages. The DPT is not very effective for this case.

Optimized logical recovery performance is on a par with SQL Server's recovery performance. Log1 redo time is practically the same as the SQL1 redo time. Log2 redo time is close to SQL2 redo time, except for the largest cache size, where it is 15% slower than SQL2 redo time. Consider for example the 512MB cache size, which is a bit more than 10% of the size of the database (a realistic value in many systems). The DPT dropped the logical redo time by 65% (from Log0 to Log1). Prefetching pages (from Log1 to Log2) dropped redo time a further 20%. So, our two optimizations (DPT and page prefetching) are very important for logical redo performance. We examine them more closely in turn.

First, the DPT can reduce the IOs due to data page stalls from 8% (for the 2048MB cache), up to 93% (for the 64MB cache). Our DPT construction scheme is very efficient. Log1 issues exactly the same data page requests as SQL1. The performance difference between them is due to the index pages needed by Log1. However, the wait time for index pages is modest, only 16% of redo time for the 64MB cache, falling to 2% of redo time for the 2048MB cache. This confirms our premise that rebuilding the database cache is the most important factor for logical recovery performance.

The overhead of constructing the DPT is the extra DC $\Delta$-log records. Figure 2(c) shows the number of $\Delta$-log records, as well as SQL Server's BW-log records seen in its analysis pass. There are more $\Delta$-log records than BW-log records because some $\Delta$-log records that contain only dirty pages need to be written as the cache fills during each checkpoint interval. The flush activity at each checkpoint cleans the cache and permits dirty page flushes to be deferred temporarily following the checkpoint. For cache sizes up to 1024MB, the number of $\Delta$-log records is no more than $1.5\times$ the number of BW-log records. We believe this is acceptable overhead to get nearly identical performance between Log1 and SQL1.

Page prefetching (described more completely in Appendix A) further reduces stalls by issuing IOs asynchronously before pages are needed. Prefetching reduces stalls for both logical and SQL Server recovery by two orders of magnitude. Running time reduction is smaller because some stalls are longer as we wait for a block of pages to be cached. With prefetching, redo performance is more variable and cannot be captured with a simple cost model. In Log2 and SQL2, data page stalls dominate if prefetching proceeds too slowly. Log page stalls dominate if prefetching fills up the cache too quickly. This is reflected in our experiments by the high variance of Log2 and SQL2 redo time. Further, when the size of the DPT is large, more pages need to be fetched, and hence prefetching is more valuable. Thus, the benefits of prefetching are much more visible in the larger cache sizes (see Figure 2(b)). Both SQL and our prefetching schemes manage to reduce the effect of the database cache size on redo performance. Our simple prefetching scheme achieves practically the same performance as SQL Server (always less than 15% slower).

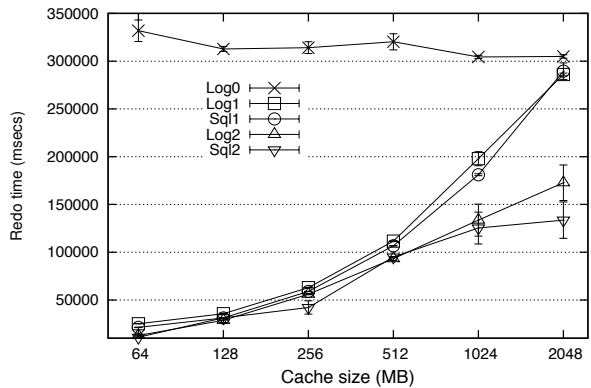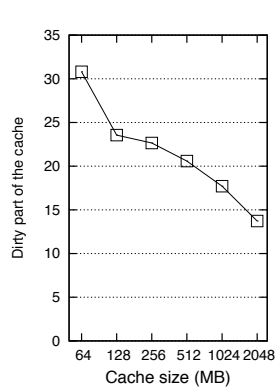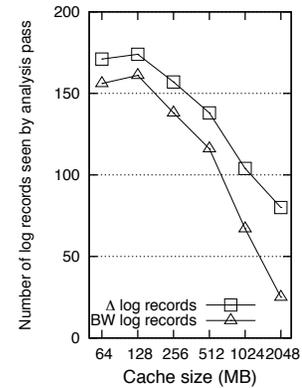

(a) Redo recovery time (msecs).  (b) Dirty percent of cache (%).  (c) $\Delta$- and BW- records written.

**Figure 2: Redo statistics for varying cache sizes.**

## 6. DISCUSSION

### 6.1 Related Work

We know of no earlier work on logical recovery. Somewhat related is recovery for client-server architectures, EXODUS [5] and ARIES/CSA [16]. A client maintains its buffer pool, the server is responsible for recovery. These papers also used a DPT, relying on the same intuition, i.e., that limiting the pages fetched during recovery is the critical factor for recovery performance.

There has been very little work done on recovery performance and most is twenty years ago or more [1, 2]. The March, 1985 issue of the Data Engineering Bulletin [8] contains a set of papers discussing recovery and its performance. Only the recent work in [17] did a side-by-side comparison of recovery techniques. Most were focused on the impact of recovery on normal execution performance, not on the performance of recovery itself in bringing a database back online. Reference [9] presents a slightly more recent simulation study of ARIES recovery performance that identifies rebuilding the database cache as the principal cost of redo recovery.

### 6.2 Conclusion

When one separates transactions from data access and knowledge of data placement, recovery needs to be expressed logically. Log records cannot contain PIDs. With a naive approach, state-of-the-art recovery optimizations cannot be used. In this paper, we have addressed the performance concerns regarding logical recovery. We have shown that with modest actions by the DC, in particular recording pages dirtied by updates to enable the construction of a DPT, these optimizations are possible. In a side-by-side comparison, our experiments demonstrate that logical recovery can achieve performance comparable to physiological recovery. We implemented one point of a spectrum of possibilities to establish this. Other points in the spectrum are discussed in Appendix D.

### ACKNOWLEDGEMENTS

Mike Zwilling invented SQL's technique for dirty page table and recovery LSNs. We benefitted greatly by conversations with Robin Dhamankar and Cristian Diaconu of the SQL team.

# APPENDIX
## A. PAGE PREFETCH

Prefetching pages avoids stalling waiting for updated pages to be read into the cache during recovery. SQL Server exploits this in its redo recovery. In addition, SQL Server can read blocks of eight contiguous pages with a single IO. We want logical recovery to be able to exploit prefetch as well.

## A.1 Prefetching Index Pages

Because logical recovery log records have no page information, all updates need to traverse the B-tree index to discover the page on which an update is to be applied. Hence, index pages are needed by the DC during recovery, even if they have never been updated.

The simplest way to acquire the needed index pages is to load them "on demand". The B-tree root will be loaded immediately, and then the path to the page on which the first redo operation is directed will be loaded. Subsequent operations will load the missing parts of their paths, and over time, searches will mostly hit pages that are already in the DC cache. However, initially, while the index pages are being loaded into the DC cache, redo will proceed very slowly as each new redo log record will need to wait for one or more index pages to be loaded before we even know what page is to be updated. Hence, even our optimized redo test needs to wait for index pages before the test can be applied.

There are several ways these index pages might be pre-loaded into the DC cache. For example, we can write DC log records that contain index page IDs, and use those to prefetch index pages during TC redo. However, at least in the common case, the number of internal index pages is very small compared to the number of data pages that need redo. Hence, we choose for the DC to simply load all index pages into memory at the beginning of DC recovery.

## A.2 Prefetching Data Pages

Prefetching data pages has a large impact on redo performance. The number of data pages that need to be brought in is typically much greater than the number of log or index pages, and the time waiting for these pages accounts for most of the redo time. Two strategies to prefetch data pages in the SQL Server setting where update log records have PIDs are log-driven and DPT-driven.

In the log-driven case, a read-ahead mechanism requests a certain number of log pages that follow the current log record. The PIDs contained in these log records are checked. If a PID is in the DPT, and the rLSN of the DPT entry is less than the LSN of the log record we are currently examining, then a prefetch for the corresponding page is issued. A disadvantage of this strategy is that a prefetch request for the same page might be issued multiple times if the same page appears in nearby log records. This is the prefetching scheme implemented in SQL Server.

DPT-driven prefetching does not depend on the log. After the DPT has been constructed, pages in the DPT are prefetched in the order of their rLSNs. This approach has the advantage of not depending on the log prefetching mechanism. Rather, data page prefetching proceeds independently. However, if there is a large gap between the first and last reference of a certain page, the page may be flushed in the meantime. In that case, redo will need to perform a synchronous IO to bring the page back. More importantly, synchronizing DPT-based prefetching with the log scan can be hard. If prefetching proceeds too quickly, pages may get flushed before the redo scan requests them. If it proceeds too slowly, redo may need to wait for the pages to be brought in.

We chose to loosely follow the SQL Server method to minimize changes. The $\Delta$-log records contain all the dirty pages in update order. During the DC analysis, we construct a list of PIDs (termed the prefetch list–PF-list) which is roughly the concatenation of the DirtySet's of $\Delta$-log records. In particular, a PID of a DirtySet is added in the PF-list if it is not already contained in the DPT. The PF-list serves as our approximation to the log. We then execute "log-driven" read-ahead using the PF-list instead of the log.

## B. PERFORMANCE ANALYSIS

Our tests confirm that redo recovery performance is mostly gated by I/O latency for data pages. Since redo starts with a clean database cache, redo performance depends on: (i) how many data pages are requested from the buffer pool (the DPT reduces the number of pages needed); and (ii) how often and how long redo waits for the pages (rLSNs and prefetching reduce waits). Log pages are fewer and are fetched sequentially and all methods process the same log. Finally, logical recovery needs to fetch index pages, a burden it does not share with physiological recovery, and this burden is included in our results. *Having the index in cache at the end of redo improves both logical undo and normal performance after redo, an advantage for logical redo that we do not measure in our results.*

Several factors impact the recovery task. First is the size of the database cache. This effect is not obvious. While a larger cache gives more room for redo to cache pages (hence fewer page swaps), during normal execution it increases the size of the DPT at recovery, and hence the number of pages that need to be fetched during recovery. We explore the effect of the database cache in Section 5.3.

Second, the distribution of the workload affects redo performance in an obvious way. The better the page locality of the workload, the fewer unique pages appear in update log records, and hence the smaller the DPT size. We use a uniform workload in our experiments, which represents the worst case for redo recovery.

Finally, the checkpoint interval affects redo time. A larger checkpoint interval implies a longer redo log and more redo log records to process. However, if the cache is in equilibrium at the time of the crash, the number of dirty pages and thus the DPT size may not be greatly affected by a larger checkpoint interval. We use the default SQL Server checkpoint interval in our workloads. In addition, we experimented with a larger checkpoint interval, and the results are as expected. We present these results in Appendix C.

When page prefetching is off (i.e., for the methods Log0, Log1, and SQL1), all IO is synchronous, and redo performance can be approximated with a simple IO model. For Log0, every page updated in the redo log since the last checkpoint will need to be fetched ("No. of log records"), in addition to the index and log pages. Assuming that every log record contains a different PID, and ignoring page swaps,

$$\text{COST}(\text{Log0}) \simeq \text{No. log records} + \text{log pages} + \text{index pages.} \quad (1)$$

For SQL1, the number of data pages that need to be brought to memory is approximately equal to the size of the constructed DPT:

$$\text{COST}(\text{SQL1}) \simeq \text{DPT size} + \text{log pages.} \quad (2)$$

Finally, for Log1, the number of data pages fetched is the size of the DPT plus the number of log records in the tail of the log. Recall that for the tail of the log recovery needs to fall back to the basic logical redo algorithm:

$$\text{COST}(\text{Log1}) \simeq \text{DPT size} + \text{No. log records in log tail} + \\ \text{log pages} + \text{index pages.} \quad (3)$$

The performance difference between Log1 and SQL1 is the burden imposed by the index pages, and more importantly, the quality of our DPT construction scheme.

Thus, the primary workload we execute (pure updates, uniformly distributed) maximizes the number of pages dirtied, and hence the number of dirty pages in the cache and therefore the size of the DPT. This is a worst case recovery scenario, even more so because we do not include reads in our workload. Reads dilute the cache "update density", meaning that fewer pages are dirty at any time.

## C. CHECKPOINT INTERVAL

The length of the checkpoint interval also impacts recovery time in our logical recovery. We varied the checkpoint interval from the default in SQL Server ($ci_1$) to $ci_2 = 5ci_1$ and $ci_3 = 10ci_1$, resulting in a $5\times$ and $10\times$ larger redone log respectively. Figure 3 shows the redo time for both the default and the enlarged intervals (in logarithmic scale). The redo time of Log0 grows linearly with checkpoint interval size. This is expected; since Log0 does not use a DPT, the number of pages that will be brought to memory is approximately equal to the number of log records (see Equation 1).

The redo times of Log1 and SQL1 are affected, but not linearly. When the checkpoint interval is 5 times larger, Log1 and SQL1 times approximately double. Two factors contribute to enlarged redo time. First, the number of log pages that need to be read is 5 times larger. Second and most importantly, the dirty part of the cache (and thus the DPT) is larger, because the checkpoint flushing activity occurs more rarely.

Log2 and SQL2 are affected only modestly by the enlarged checkpoint interval. Their redo times are 1.2 times larger than Log1 and SQL1 respectively when the checkpoint interval grows from $ci_1$ to $ci_2$. With the enlarged checkpoint interval, there are 5 times more log records (and thus 5 times more calls to the data page prefetching routine), but only twice as many pages to be prefetched (the size of the DPT). Thus, the benefits of prefetching are more visible in this case. The same effect can be observed when the checkpoint interval grows from $ci_2$ to $ci_3$. The effect of the checkpoint interval is the same for both our logical, and SQL Server recovery.

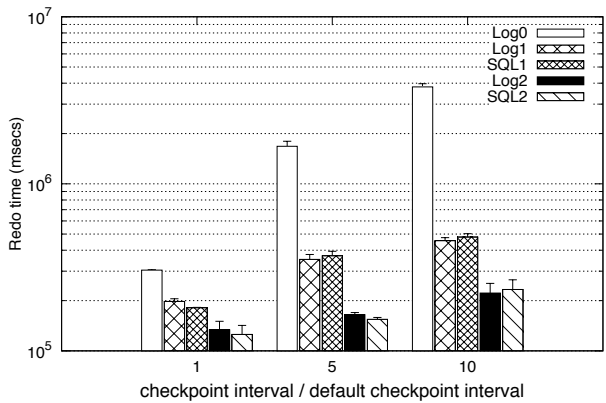

**Figure 3: Redo time (msecs, logarithmic scale) when varying the checkpoint interval from the SQL Server default ($ci_1$) to $ci_2 = 5ci_1$ and $ci_3 = 10ci_1$.**

## D. SOME ALTERNATIVES

The DPT construction algorithm described in Section 4.2 is not the only one that can be followed by logical recovery. It is merely one point in a spectrum of choices, representing a particular trade-off between the amount of DC logging and the accuracy of the final DPT. Our point on this spectrum is toward the low logging end to keep overhead during normal operation low. Indeed, we log roughly as much as SQL Server does for integrated recovery. At the same time, the constructed DPT has roughly the same accuracy as the DPT constructed by SQL Server. Alternative choices we explored are discussed in Section D.

### D.1 "Perfect" DPT

The DC can construct an almost perfect DPT (excluding the log tail), if it captures every update to a page together with the update's LSN. This requires adding an array of LSNs to be kept at runtime, and added to the $\Delta$-log record, call it the DirtyLSNs array. This means that $\Delta$-log records will be written more frequently, or that they will occupy more space. At the time of the DC analysis pass, the DC has enough information to construct exactly the same DPT as SQL Server. In theory, this means that logical recovery will have exactly the same performance as traditional recovery (excluding the overhead of traversing the B-tree, and handling the tail of the log). Although this choice is attractive, it requires additional logging compared to SQL Server recovery.

### D.2 Reduced Logging

At the other end of the spectrum, we can exclude the FW-LSN and FirstDirty fields from the $\Delta$-log record. In this case, all the pages in the dirty array, when added to the DPT during analysis, must have an rLSN set to the TC-LSN of the previous $\Delta$-log record. The flushed set of a $\Delta$-log record can be used to prune pages from the DPT that were added when prior $\Delta$-log records were processed, but not from the current $\Delta$-log record. We chose to add the two fields to our $\Delta$-log record because the logging cost is minimal, and this information improves recovery performance somewhat.

When monitoring dirty and flushed pages, we might strive to avoid duplicates. For example, if a page is dirtied and then cleaned, we may want to remove its PID from the dirty PID array. However, this would require deletes in the dirty PID array, which are order N operations. We decided to avoid that to reduce overhead during normal operation.